\documentclass[epj,nopacs,fleqn]{svjour}
\pdfoutput=1
\usepackage{slashed,graphicx,amsmath,amssymb,cite,url,hyperref,
  subfigure,graphicx,placeins,xspace}
\usepackage[utf8]{inputenc}
\hypersetup{bookmarks=true,unicode=true,pdftoolbar=true,pdfmenubar=true,
  pdffitwindow=false,pdfstartview={FitH},
  pdfnewwindow=true,colorlinks=true,
  linkcolor=blue,citecolor=magenta,filecolor=magenta,urlcolor=cyan}

\usepackage[absolute]{textpos}

\def\muone{$m_{\tilde{u}_1}$}
\def\ifb{{\rm fb}^{-1}}

\def\etmiss{\ensuremath{E_{\mathrm{T}}^{\mathrm{miss}}}\xspace}
\def\ptmiss{\ensuremath{\vec p^{\mathrm{\ miss}}_\mathrm{T}\xspace}}
\def\ptl{\ensuremath{\vec p^{\mathrm{\ \ell}}_\mathrm{T}\xspace}}
\newcommand{\mttwo}{\ensuremath{m_{\mathrm{T2}}}\xspace}
\newcommand{\mttwoblj}{\ensuremath{m_{\mathrm{T2}_{blj}}}\xspace}
\newcommand{\amttwo}{\ensuremath{am_{\mathrm{T2}}}\xspace}
\newcommand{\pt}{\ensuremath{p_{\mathrm{T}}}\xspace}
\newcommand{\ttbarW}{\ensuremath{t\bar{t}W}\xspace}
\newcommand{\ttbarZ}{\ensuremath{t\bar{t}Z}\xspace}
\newcommand{\mTlep}{\ensuremath{m_\mathrm{T}^{lep}}\xspace}
\newcommand{\ttbar}{\ensuremath{t\bar{t}}\xspace}
\newcommand{\Wjets}{\ensuremath{W}+jets\xspace}
\newcommand{\Zjets}{\ensuremath{Z}+jets\xspace}
\newcommand{\btagged}{\ensuremath{b}-tagged\xspace}
\def\GeV{\ifmmode {\mathrm{\ Ge\kern -0.1em V}}\else
                   \textrm{Ge\kern -0.1em V}\fi}%

\begin{document}

\title{Flavour-violating decays of mixed top-charm squarks at the LHC}

\begin{textblock}{5}(10.7,0.75) 
	LAPTH-024/18, KEK-TH-2072
\end{textblock}

\author{
  Amit Chakraborty\inst{1}\thanks{\color{blue}amit@post.kek.jp},
  Motoi Endo\inst{1}\thanks{\color{blue}motoi.endo@kek.jp},
  Benjamin Fuks\inst{2,3}\thanks{\color{blue}fuks@lpthe.jussieu.fr},
  Bj\"orn Herrmann\inst{4}\thanks{\color{blue}herrmann@lapth.cnrs.fr},
  Mihoko M. Nojiri\inst{1,5,6}\thanks{\color{blue}nojiri@post.kek.jp},
  Priscilla Pani\inst{7}\thanks{\color{blue}priscilla.pani@cern.ch} and
  Giacomo Polesello\inst{8}\thanks{\color{blue}giacomo.polesello@cern.ch}
}

\institute{
  Theory Center, IPNS, KEK, Tsukuba, Ibaraki 305-0801, Japan
  \and
  Sorbonne Universit\'e, CNRS, Laboratoire de Physique Th\'eorique et Hautes
    Energies, LPTHE, F-75005 Paris, France
  \and
  Institut Universitaire de France, 103 boulevard Saint-Michel, 75005 Paris,
    France
  \and
  Univ.\ Grenoble Alpes, USMB, CNRS, LAPTh, F-74000 Annecy, France
  \and
  The Graduate University of Advanced Studies (Sokendai), Tsukuba, Ibaraki
    305-0801, Japan
  \and 
  Kavli IPMU (WPI), University of Tokyo, Kashiwa, Chiba 277-8583, Japan 
  \and
  CERN, Experimental Physics Department, CH-1211 Geneva 23, Switzerland
  \and
  INFN, Sezione di Pavia, Via Bassi 6, 27100 Pavia, Italy
}

\date{}

\abstract{
We explore signatures related to squark decays in the framework of non-minimally flavour-violating Supersymmetry. We consider a simplified model where the lightest squark consists of an admixture of charm and top flavour. By recasting the existing LHC searches for top and charm squarks, we show that the limits on squark masses from these analyses are significantly weakened when the top-charm mixing is sizeable. We propose a dedicated search for squarks based on the $tc+{E_{\mathrm{T}}^{\mathrm{miss}}}$ final state which enhances the experimental sensitivity for the case of high mixing, and we map its expected reach for the forthcoming runs of the LHC. We emphasize the role of analyses requiring a jet tagged as produced by the fragmentation of a charm quark in understanding the squark mixing pattern, thus providing a novel handle on new physics. Our results show that, in order to achieve full coverage of the parameter space of supersymmetric models, it is necessary to extend current experimental search programmes with analyses specifically targeting the cases where the lightest top-partner is a mixed state.
}

\titlerunning{Flavour-violating decays of mixed top-charm squarks at the LHC}
\authorrunning{A.~Chakraborty {\it et al.}}

\maketitle
\flushbottom

\section{Introduction}
\label{sec:intro}

While the Large Hadron Collider (LHC) pursues its quest for new physics, its
Run~2 at a centre-of-momentum energy of 13 TeV being on-going, no direct
evidence for physics beyond the Standard Model (SM) has been observed so far.
The conceptual problems and limitations of the Standard Model therefore remain
unsolved. Among the various theoretical frameworks tackling
those issues, Supersymmetry still remains one of the most popular and appealing
options. The absence of experimental evidence for any of the
superpartners of the Standard Model degrees of freedom, and in particular of
the strongly-interacting squarks and gluinos \cite{Sirunyan:2017kqq,
Aaboud:2017dmy,Sirunyan:2017xse,Sirunyan:2017wif,Sirunyan:2017kiw,
Aaboud:2017nfd,Aaboud:2017wqg,Aaboud:2017ayj,Sirunyan:2017pjw,Sirunyan:2017leh,
Aaboud:2017phn,Aaboud:2017aeu,Aaboud:2017vwy, Sirunyan:2017cwe,Aaboud:2018zjf},
however imposes strong constraints on how
Supersymmetry could be viably realised. For instance, the superpartners have to
be too heavy to be produced at current LHC centre-of-mass energy and luminosity,
or the particle spectrum has to be highly degenerate~\cite{%
Martin:2007gf, Fan:2011yu, Murayama:2012jh}. An alternative to
these two solutions would be to abandon the idea of a minimal supersymmetric
realisation which most current searches are based upon.

The ATLAS and CMS collaborations have obtained their limits
under the assumption of a rather simplified realisation of the Minimal
Supersymmetric Standard Model (MSSM). Those studies generally consider that a
limited number of new physics states are light and can thus be produced at the LHC, and
that those light particles therefore undergo a single well-defined decay mode.
While studies of such simplified situations
are reasonable and important starting points, the structure of the MSSM could be
more general and complex and yield decay patterns that are 
not addressed by current searches.

In the present work, we address this last point and consider a realisation of
the MSSM where inter-generational mixings in the squark sector are allowed. In
this case, a physical squark eigenstate contains several components of
well-defined flavour, which opens the door to multiple decay modes. For
example, if the lightest squark is dominantly of top flavour but
additionally contains a sizeable charm-flavour component, decays involving a
top quark and a charm quark may both have
a significant branching fraction. As a consequence, not all
signatures stemming from squark pair production and decay are captured by
the existing experimental searches, and current direct search limits 
may be weakened~\cite{Blanke:2013zxo,Brooijmans:2018xbu}.

In the MSSM realisations usually under consideration, squark
inter-ge\-ne\-ra\-ti\-o\-nal
mixings are suppressed by imposing the Minimal Flavour Violation (MFV) paradigm
in which all flavour-violating interactions originate from the diagonalisation
of the fermion sector and the corresponding CKM and PMNS matrices. Departing
from an MFV squark sector, additional flavour-mixing terms may be present in
the Lagrangian of the theory. By virtue of kaon data constraints, any mixing
involving squarks of the first generation has to be extremely
small~\cite{Ciuchini:2007ha}, although second and third generation squark
mixings are still largely allowed by current data~\cite{DeCausmaecker:2015yca}.
We therefore focus on mixings solely involving the charm and top flavours of
right-handed squarks. Such mixings can for
instance originate from Grand Unification at a high scale~\cite{Dimou:2015yng,
Dimou:2015cmw}, and their TeV scale implications on squark
production and decay have received considerable attention in the
past~\cite{Bozzi:2007me,Fuks:2008ab,Hurth:2009ke,Bruhnke:2010rh,Bartl:2010du,
Bartl:2011wq,Bartl:2012tx,Backovic:2015rwa,Blanke:2015ulx,Crivellin:2016rdu,%
Blanke:2017tnb,Blanke:2017fum,Evans:2015swa}.
In this work, we assess how they are constrained by the most recent LHC results
and show, by proposing a novel class of search strategies, how future LHC
searches could be tailored better to constrain non-minimally flavour-violating
(NMFV) Supersymmetry.

The rest of this paper is organised as follows. In Sec.~\ref{sec:model}, we
introduce the simplified setup that we have adopted to include
inter-generational squark mixings in the MSSM. We then reinterpret the results
of
existing LHC direct searches for squarks when potentially altered squark decay
modes are allowed. In Sec.~\ref{sec:collider}, we present an analysis strategy
targeting the specific channel where, after squark pair-production, 
one squark decays into a top
quark and the second squark into a charm quark. Projections for 
sensitivity of the coming LHC runs to the model are presented and discussed in
Sec.~\ref{sec:results}, where we emphasise the potential role of charm tagging.
Our conclusions are given in Sec.~\ref{sec:summary}.

\section{Model setup and exisiting LHC limits}
\label{sec:model}

In this section, we describe the simplified model featuring top-charm mixing in
the squark sector which we base our analysis on. We discuss general features
of the search strategy for targeting such a model, and reinterpret
recent squark searches to assess the current LHC coverage of the model parameter
space.


\subsection{A simplified model for squark flavour violation}
\label{lhcnmfv_SecModel}

In order to assess the LHC sensitivity to supersymmetric models featuring non-minimal flavour violation
in the squark sector, we consider a simplified model embedding two active
flavours of squarks, a right-handed top squark $\tilde{t}_R$ and a right-handed
charm squark $\tilde{c}_R$. These two states mix into two physical eigenstates
$\tilde{u}_1$ and $\tilde{u}_2$ whose flavour structure is dictated by the
$\theta_{tc}$ mixing angle,
\begin{equation}
	\begin{pmatrix} \tilde{u}_1 \\ \tilde{u}_2 \end{pmatrix} =
	\begin{pmatrix} ~~\cos\theta_{tc} & ~\sin\theta_{tc} \\ -\sin\theta_{tc} & ~\cos\theta_{tc} \end{pmatrix}
	\begin{pmatrix} \tilde{c}_R \\ \tilde{t}_R \end{pmatrix} \ ,
\end{equation}
where by convention $\tilde{u}_1$ is the lighter of the two mass eigenstates.
Our simplified model additionally includes one neutralino $\tilde{\chi}^0_1$,
that we take bino-like. 
Such an assumption does not 
have a significant impact on our phenomenological results.
The setup of our interest is thus based on four parameters: the masses $m_{\tilde{u}_1}$ and $m_{\tilde{u}_2}$ of the two physical squarks together with the flavour mixing angle $\theta_{tc}$, and the neutralino mass $m_{\chi^0_1}$. 
For the sensitivity studies in the $(m_{\tilde{u}_1}, \theta_{tc})$ plane,
the neutralino mass will be fixed to $m_{\chi^0_1} = 50$ GeV.
Although a more complicated flavour
structure involving left-handed squarks could be possible as well, this last
setup implies the need to handle more complicated constraints from $B$-physics in order to build
phenomenologically viable scenarios. Left-handed squarks are thus assumed
heavier and decoupled, like any other superpartner.

Our simplified model therefore exhibits two competing squark decay modes (if
kinematically allowed),
\begin{equation}
	\tilde{u}_i \to t \tilde{\chi}^0_1 \,, \qquad \tilde{u}_i \to c \tilde{\chi}^0_1 \qquad\text{with}\ \ i=1,2\ ,
  \label{sqdecays}
\end{equation}
which yield three classes of LHC signatures originating from the production of a
pair of $\tilde{u}_i$ squarks. 
Typical LHC search strategies have been designed on the basis of the 
MFV pa\-ra\-digm and thus only
address the two signatures:
\begin{equation}
  pp \to t\bar{t} + \etmiss \qquad\text{and}\qquad
  pp \to c\bar{c} + \etmiss \,,
  \label{decays}
\end{equation}
where $\etmiss$ is the imbalance in transverse momentum 
in the event generated by the undetected neutralinos.
Squark flavour mixing opens up a third final state,
\begin{equation}
	pp \to tc + \etmiss \,,
	\label{nmfvdecay}
\end{equation}
where one squark decays into a top quark and the other one into a charm
quark~\cite{Bartl:2010du}. 

\begin{figure*}
	\begin{center}
	\includegraphics[width=.495\textwidth]{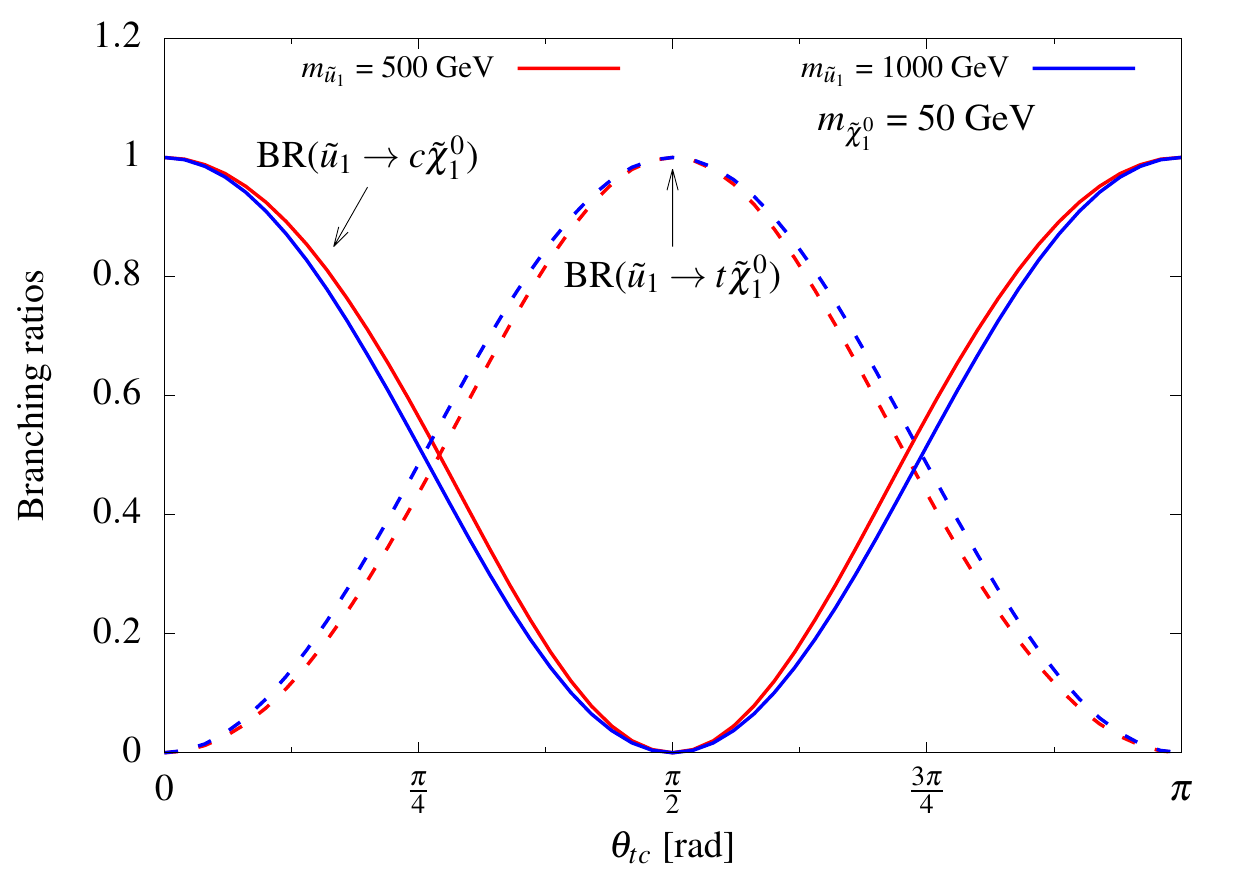}
	\includegraphics[width=.495\textwidth]{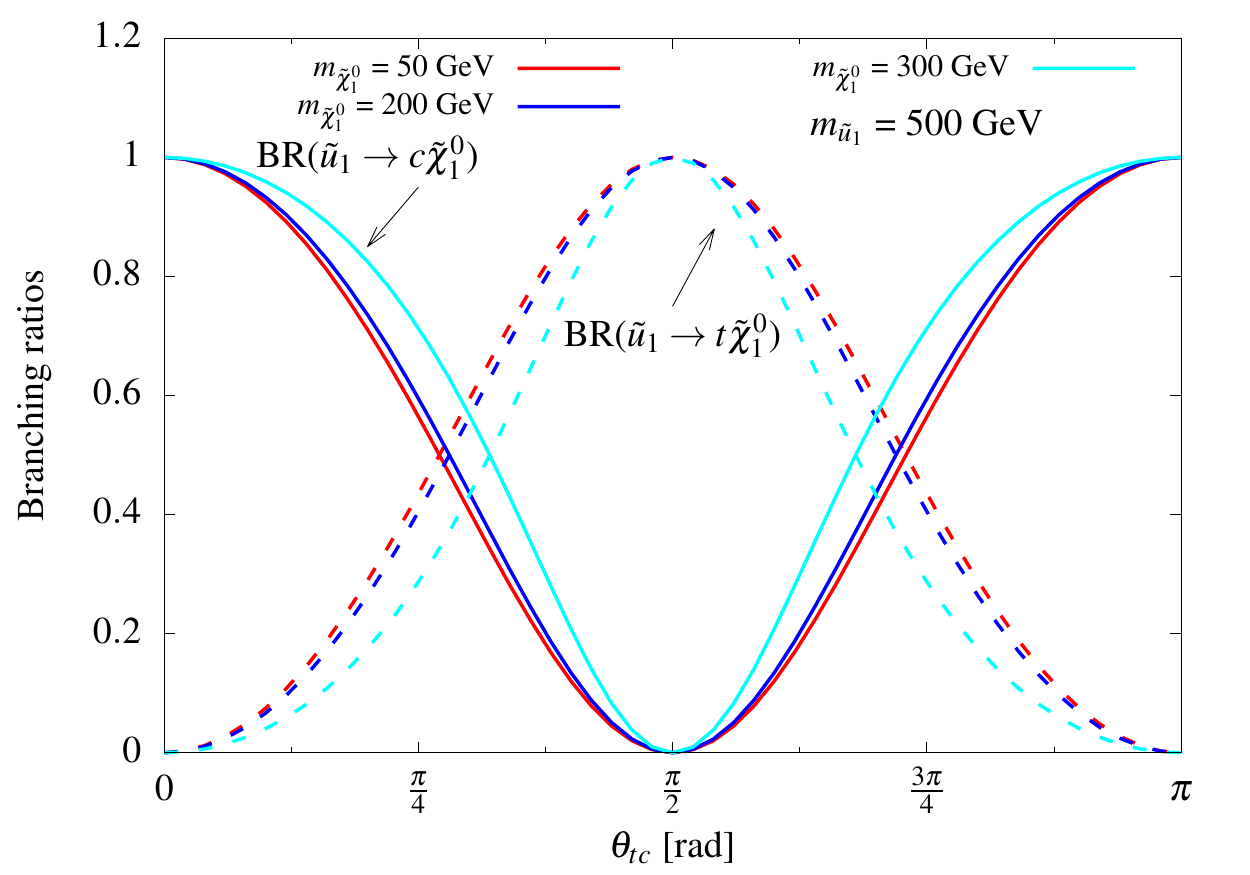}
	\end{center}
	\vspace*{-5mm}
	\caption{Dependance of the branching ratios
   BR($\tilde{u}_1\to t\tilde{\chi}^0_1$) (dashed) and
   BR($\tilde{u}_1\to c\tilde{\chi}^0_1$) (solid) on the squark mixing angle
   $\theta_{tc}$ for various mass configurations. In the left panel, the squark
   mass is fixed to 500~GeV (red) and 1000~GeV (blue), with the neutralino mass
   being set to $m_{\chi^0_1} = 50$~GeV. In the right panel, the neutralino mass
   varies and is fixed to 50~GeV (red), 200~GeV (blue) and 300~GeV (cyan), for a
   squark mass of $m_{\tilde{u}_1}=500$~GeV.}
	\label{lhcnmfv_fig:bratio}
\end{figure*}

In Fig.~\ref{lhcnmfv_fig:bratio}, we illustrate the $\theta_{tc}$-dependence of
the squark branching ratios associated with the decays of
Eq.~\eqref{sqdecays}. We observe that regardless the squark and neutralino
mass configuration, there always exists a $\theta_{tc}$ value for which 
both decay modes have 50\% branching ratio, 
which means that half of the signal events 
would produce the final state of Eq.~\eqref{nmfvdecay}. 
Moreover, differences in the functional
behaviour of the branching ratios for different mass hierarchies
become only noticeable close to threshold,
when the mass splitting between the decaying squark and the neutralino is
small. This configuration is not considered further in this paper, as 
the phase space available for the decay is limited and 
the best experimental sensitivity is achieved with
monojet or monotop probes~\cite{Fuks:2014lva}.

\subsection{Reinterpretation of current LHC Run~2 results}
\label{sec:recast}
\begin{figure*}
	\begin{center}
	\includegraphics[width=.476\textwidth]{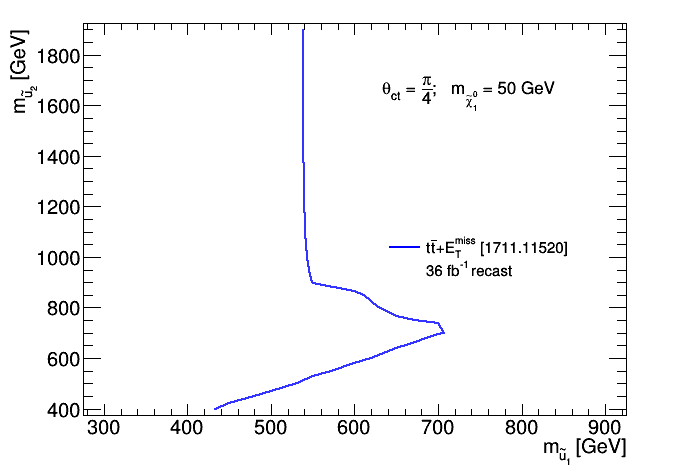}
	\includegraphics[width=.495\textwidth]{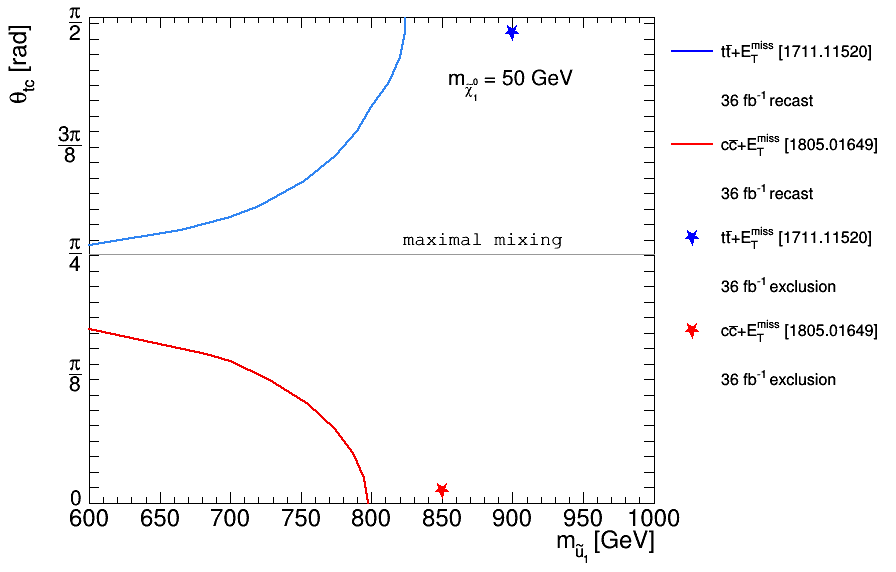}
	\end{center}
	\vspace*{-5mm}
	\caption{Reinterpretation, in the context of our simplified model, of
  the ATLAS bounds on top squarks obtained with single-leptonic probes~\cite{
  Aaboud:2017aeu} and on Supersymmetry when charm tagging is used~\cite{
  Aaboud:2018zjf}. The results are presented in the $(m_{\tilde{u}_1},
  m_{\tilde{u}_2})$ plane (left) and $(m_{\tilde{u}_1}, \theta_{tc})$ plane
  (right), and the stars correspond to the official ATLAS results in the
   non-flavour-mixing case. The excluded region lies between the exclusion 
contour and the left-side of the figure.
}
	\label{lhcnmfv_fig:recast}
\end{figure*}

The ATLAS and CMS collaborations have performed several direct searches for top
squarks, mostly in a setup where they are pair-produced and decay into a pair of
top-antitop quarks and missing energy as indicated by the first equation of
Eq.~\eqref{decays}. With no observation of any hint for new physics, the most
stringent constraints arise from LHC Run~2 analyses of proton-proton collisions
at a centre-of-mass energy of 13 TeV \cite{Sirunyan:2017kqq,Aaboud:2017dmy,
Sirunyan:2017xse,Sirunyan:2017wif,Sirunyan:2017kiw,Aaboud:2017nfd,
Aaboud:2017wqg,Aaboud:2017ayj,Sirunyan:2017pjw,Sirunyan:2017leh,Aaboud:2017phn,
Aaboud:2017aeu}. All these searches lead to exclusion limits on the top
squark mass of the order of 1~TeV. Bounds on first and second generation squarks
are similar when one single light squark species is considered together
with a decay into light jets and missing transverse energy, whereas they
reach 1.5~TeV for models featuring four mass-degenerate first and second
generation squarks~\cite{Aaboud:2017vwy,Sirunyan:2017cwe}. 
The most sensitive stop
searches, yielding a similar expected sensitivity for low neutralino masses, are
the ones addressing final states with either zero or one lepton. We therefore
choose the recent ATLAS search for top squarks in final states with one lepton of
Ref.~\cite{Aaboud:2017aeu} as a benchmark for getting conservative Run~2
constraints on our model.\par
Additionally, the ATLAS collaboration has carried out an analysis
targeting top squarks decaying into
charm and missing energy or charm squarks~\cite{Aaboud:2018zjf},
based on the experimental tagging of jets produced from the 
fragmentation of charm quarks.
As this signature is expected to
play a significant role for getting handles on the considered squark
inter-generational mixings, we use the analysis in Ref.~\cite{Aaboud:2018zjf} as a second LHC
Run~2 benchmark to evaluate the existing constraints on our simplified model.\par
We perform a three-dimensional parameter space scan and vary independently the
two squark masses ($m_{\tilde{u}_1}$ and $m_{\tilde{u}_2}$), as well as the
top-charm squark mixing angle $\theta_{tc}$. As mentioned above, the neutralino
mass has been fixed to 50~GeV, so that our results are valid as long as the
squark masses are much larger than the neutralino mass. For each considered
point, we
evaluate the sensitivity of the two searches of Refs.~\cite{Aaboud:2017aeu,
Aaboud:2018zjf} and present the results in Fig.\ \ref{lhcnmfv_fig:recast}. 
The excluded region lies between the exclusion
contour and the left-side of the figure. Concerning the stop analysis~\cite{Aaboud:2017aeu}, we rely on the
acceptances and efficiencies that have been officially provided by the ATLAS
collaboration for each of the `discovery tN\_med' (targeting moderate stop
masses) and `discovery tN\_high' (targeting high stop masses) regions. We then
estimate the two corresponding signal yields ($N_{\rm sig}$), considering
next-to-leading order (NLO) stop pair-pro\-duc\-ti\-on ra\-tes corrected by the
resummation of the threshold logarithms at the next-to-leading logarithmic
(NLL) accuracy~\cite{Borschensky:2014cia} and the appropriate
branching ratios. These signal yields are then compared to the ATLAS
model-independent upper limit ($N^{\rm obs~limit}_{\rm non-SM}$) for each of the
regions. If the ratio of these two yields exceeds one, the signal point is
considered excluded. While providing acceptance and efficiency values 
only for the inclusive `signal' regions,  the ATLAS
analysis employs a multi-bin fit in the most sensitive distribution 
for the final exclusion limit estimation.
For this reason the recast exclusion contours presented in
Fig.~\ref{lhcnmfv_fig:recast} represent a conservative estimate of the 
effective reach of the ATLAS search. We rely on the same procedure to extract the constraints from
the charm-tagging analysis of Ref.~\cite{Aaboud:2018zjf}.

In the left panel of Fig.~\ref{lhcnmfv_fig:recast}, we consider a class of
benchmark scenarios
where the two squark eigenstates are maximal admixtures of the top and charm
flavours ($\theta_{ct} = \frac \pi 4$) and we vary the two masses independently (with $m_{\tilde{u}_1}<
m_{\tilde{u}_2}$). The total new physics production rate is here solely driven
by the lightest of the two states, except for the region where the mass
splitting of the two squarks is small. For sufficiently high splittings, the exclusion is thus
independent of $m_{\tilde{u}_2}$, and squarks are found to be constrained
to be heavier than about 550~GeV. Compared with the more standard MFV case where
the two eigenstates are also flavour eigenstates (and where the bounds are of
about 1~TeV), the limits are hence weakened by almost 500~GeV.
The large value of the top-charm mixing angle indeed implies that the two signal
regions of the stop analysis of Ref.~\cite{Aaboud:2017aeu}, specifically
targeting final states with the decay products of two top quarks, 
are less populated by virtue of the
large decay fraction into charm jets BR$(\tilde{u}_1 \to c \tilde{\chi}^0_1)$.
In the parameter space region defined by
\begin{equation}
  m_{\tilde{u}_1},  m_{\tilde{u}_2} \lesssim 750~{\rm GeV} \ ,
\end{equation}
the situation is somewhat different as the two squark mass eigenstates
contribute to a potentially observable new phy\-sics signal. This partly
compensates the loss due to the smaller branching ratio into tops, so that the
obtained limits are  stronger than when the second eigenstates is
heavier. The charm-tagging analysis of Ref.~\cite{Aaboud:2018zjf} always implies
weaker bounds for this specific classes of scenarios (the number of events
populating the signal regions being very small), and the corresponding results
are thus omitted.

In the right panel of Fig.~\ref{lhcnmfv_fig:recast}, we 
reinterpret the ATLAS limits in the $(m_{\tilde u_1}, \theta_{tc})$ plane, 
{\it i.e.} we decouple the second eigenstate. Our
results exhibit the complementary effect of the top-charm squark mixing angle on
the bounds. For $\theta_{tc}=0$, the lightest squark is purely of charm
flavour, so that the ATLAS stop search is insensitive to the signal and the
limits ($m_{\tilde u_1} \gtrsim 800$~GeV) solely arise from the ATLAS
charm-tagging analysis. With the mixing angle increasing, the $c\bar{c}+\etmiss$
production rate decreases so that the bounds are progressively weakened.
On the other
hand, the increase in $\theta_{tc}$ implies that while the signal regions of the
charm-tagging analysis are more and more depleted due to the lower and lower
BR$(\tilde{u}_1 \to c\tilde{\chi}^0_1)$ branching ratio, the signal regions of
the stop analysis are more and more populated due to the increasing
BR$(\tilde{u}_1 \to t \tilde{\chi}^0_1)$ branching ratio. 
In the
limit for which the lightest squark is purely of top flavour, its mass is
constrained to be at least  825~GeV. In the maximal-mixing condition, the mass constraints for both analyses are 
below 600~GeV, which is the minimum mass value for which experimental
acceptances are available for both the considered benchmark analyses.

We superimpose to the  results of our recasting the official
limits observed by ATLAS, represented by stars on the right panel of
Fig.~\ref{lhcnmfv_fig:recast} for the cases where the lightest squark is of a
definite flavour. The usage of multi-bin signal regions 
increases the limits by about 50--100~GeV.

\section{Collider projections for the reach of the $tc$ channel} 
\label{sec:collider}
We have shown in the previous section that the current experimental searches 
focusing on pair production of squarks 
that carry a well-defined flavour have a significantly reduced sensitivity
to models with sizeable values of flavour mixing. 
The issue may be addressed by developing a dedicated analysis 
targeting the $tc + \etmiss$ channel, which has its maximum rate
in the case of maximum mixing. We describe a possible implementation
of such an analysis in this section.
In particular, we will focus on the case in which the top quark
decays semileptonically, resulting in a final state with 
an isolated lepton, a $b$-jet, a $c$-jet and  missing transverse energy.\par
Our study assumes proton-proton collisions at a centre-of-mass energy 
of 14~TeV, and integrated luminosities of 300 and 3000~$\ifb$,
corresponding to the expected configurations for the coming LHC runs.

\subsection{Monte Carlo simulation}

In order to simulate our signal, we have implemented the model of Sec.~\ref{%
lhcnmfv_SecModel} into {\sc Feynrules 2.0}~\cite{Alloul:2013bka} to get a UFO
model~\cite{Degrande:2011ua} to be used within the
{\sc MadGraph5}\_aMC@NLO framework~\cite{Alwall:2014hca}. We have generated
leading-order (LO) hard-scattering matrix elements for squark pair-production
and decay, that we have convoluted with the leading-order set of NNPDF~3.0
parton
distribution functions~\cite{Ball:2014uwa}. Parton showering and hadronisation
have been handled with {\sc Pythia 8.2}~\cite{Sjostrand:2014zea}, and each event
has been reweighted so that the corresponding total rate matches the production
cross-section estimated at the NLO+NLL accuracy~\cite{Borschensky:2014cia}.
We generate a grid in the parameter space of the model, the lightest squark
mass being varied in the [600~GeV, 1.5~TeV] window by steps of 100~GeV, 
and the neutralino mass in the [50~GeV, 900~GeV] window in steps 
of 50~GeV for $m_{\tilde\chi^0_1}<400\, \GeV$ and of 100~GeV above.
The squark mixing angle is fixed  to $\pi/4$.

As stated above, we focus on the signal topology with one isolated 
lepton (electron or muon), jets and missing transverse energy.
The SM processes which can mimic this topology 
involve one or two leptons originating either from the 
decay of a $W$ or a $Z$ boson, or from
leptonically-decaying tau leptons. We consequently generate events for SM
\ttbar, $Wt$,  $t$-channel single top, \ttbarW, \ttbarZ, $tWZ$, $tZ$, \Wjets,
\Zjets, $WW$, $WZ$  and $ZZ$ production. For \ttbar, single top and diboson
processes, events are simulated at the NLO in QCD within the
{\sc Powheg Box} framework \cite{Alioli:2010xd}. Samples for the remaining
processes are then generated at LO, using {\sc MadGraph5}\_aMC@NLO. We consider
matrix elements featuring a variable number of additional jets that we merge
according to the CKKW prescription as implemented in {\sc Pythia
8}~\cite{Lonnblad:2011xx}. For \Wjets and \Zjets, we merge samples
describing final-states containing up to four additional partons, whereas for
\ttbarW and \ttbarZ production, the matrix elements are allowed to include up to
two extra partons. All those events are reweighted so that the total rates match
the next-to-next-to-leading order (NNLO) cross-sections if available, or the
NLO ones otherwise.\par
Jets are reconstructed according to the anti-$k_T$ jet algorithm~\cite{%
Cacciari:2008gp} with a jet radius parameter set to $R = 0.4$, as implemented in
{\sc FastJet}~\cite{Cacciari:2011ma}. Moreover, jets are labelled as $b$-jets
if the angular distance  $\Delta R\equiv(\Delta\phi^2+\Delta\eta^2)^{1/2}$ 
between the jet and the nearest
$B$-hadron satisfies $\Delta R < 0.5$. Similarly, we define $c$-jets as jets
that fail $b$-tagging and for which there exists a charmed hadron lying at an
angular distance smaller than 0.5 from the jet. Any jet that is not identified
as a $b$-jet or as a $c$-jet is labelled as a light jet.
The missing transverse momentum $\ptmiss$, with magnitude $\etmiss$,
is estimated by the vector sum of the 
transverse momenta of all invisible particles.\par
Detector effects are simulated by smearing the momenta of all reconstructed
objects and by applying reconstruction efficiency factors in a way that
reproduces the performance of the ATLAS detector~\cite{Aad:2008zzm, Aad:2009wy},
as described in Ref.~\cite{Pani:2017qyd}. 
In particular, we include $b$-tagging and $c$-tagging efficiency and rejection
factors based on the performance reported in Refs.~\cite{ATL-PHYS-PUB-2015-022,
ATL-PHYS-PUB-2017-013}, and we adopt working points corresponding to an
average $b$-tagging efficiency of $\epsilon_b(b)=77\%$ for charm and light jet
rejection factors of $1/\epsilon_b(c)=4.5$ and $1/\epsilon_b(l)=140$
respectively, and an average $c$-tagging efficiency of $\epsilon_c(c)=30\%$ for
rejection factors of $1/\epsilon_c(b)=18$ and $1/\epsilon_c(l)=5$ for $b$-jets
and light jets respectively. Such a choice is aimed at optimising background
rejection when it is dominated by final states featuring two $b$-jets. As there
is currently no public information on the correlations between the $b$-tagging
and $c$-tagging algorithms used by the collaborations, we do not allow a jet to
be $b$-tagged and $c$-tagged simultaneously. We instead first select jet
candidates based on their kinematics before applying either $b$-tagging or
$c$-tagging.\par
We have compared our approach with an independent simulation based
on the publicly available detector simulation software {\sc Delphes 3}~%
\cite{deFavereau:2013fsa}, and have found good agreement between the two
methods.

\subsection{Event selection}
\label{sec:anal1}

The topology of interest includes one
isolated lepton (electron or muon) arising from the top decay, jets including one $b$-jet
(also issued from the top decay) and one $c$-jet, as well as missing transverse
energy carried by the two neutralinos. Consequently, we preselect events
by requiring the presence of exactly one isolated electron or muon with a
transverse momentum \mbox{$\pt>25$~GeV} and a pseudorapidity $|\eta|<2.5$, and
of at least one \btagged jet with $\pt>50\, \GeV$ and $|\eta|<2.5$. We moreover
ask the invariant mass of at least one of the possible systems made of a $b$-jet
and the lepton to
fulfil $m_{b\ell} < 160\, \GeV$, since in the signal case the lepton and the
$b$-jet originate from a top decay so that $m_{b\ell}$ is bounded to be smaller
than around 153~GeV.

The dominant backgrounds at this point are comprised of $\ttbar$ events with
either one or both top quarks decaying leptonically, single top events and
$\ttbarZ$ events with an invisible $Z$-boson decay. As all backgrounds where the
missing energy originates from a leptonic $W$-boson decay ($W \to \ell \nu$)
feature
\begin{equation}
  \mTlep \!\equiv\!\sqrt{2\,|\ptl|\,\etmiss\,(1\!-\!\cos\Delta\phi(\ptl,\ptmiss))}<
  m_W\ ,
\end{equation}
we require $\mTlep>160$~GeV to increase the signal over background ratio. In
the definition of $\mTlep$,  
$\ptl$ is the lepton (vector) transverse momentum and $\Delta\phi(\ptl,\ptmiss)$ the  angle
between $\ptl$ and $\ptmiss$. Moreover, most of these backgrounds exhibit two
$b$-jets in the final state, whereas the signal features in contrast one $b$-jet
and one $c$-jet. Two strategies can therefore be envisaged to separate the
signal from the backgrounds. Either one could veto the presence of any
additional $b$-tagged jet besides the one required at the preselection level
(Case-A), or one could enforce, in addition, the presence of an extra $c$-tagged
jet (Case-B). From naive calculations based on the efficiencies of the different
tagging algorithms, the signal over background ratio is improved by a factor of
about 1.5 more for the Case-B strategy, but at the price of an overall reduction
in statistics by a factor of approximately 3. Both approaches are thus pursued
in the following. For the Case-A strategy, we veto the presence of any extra
$b$-jet and impose that there is an extra light jet with $\pt>100\, \GeV$ failing $b$-tagging. In
contrast, for the Case-B strategy, we require that only one $b$-tagged jet
satisfies $m_{b\ell} < 160\, \GeV$, we additionally impose that the leading jet
fullfilling $m_{j\ell} > 160\, \GeV$ is $c$-tagged and has a $\pt>100$~GeV, and
we ask all remaining jets with $m_{j\ell} > 160\, \GeV$ to fail $b$-tagging.

\begin{figure*}
	\begin{center}
	\includegraphics[width=0.495\textwidth]{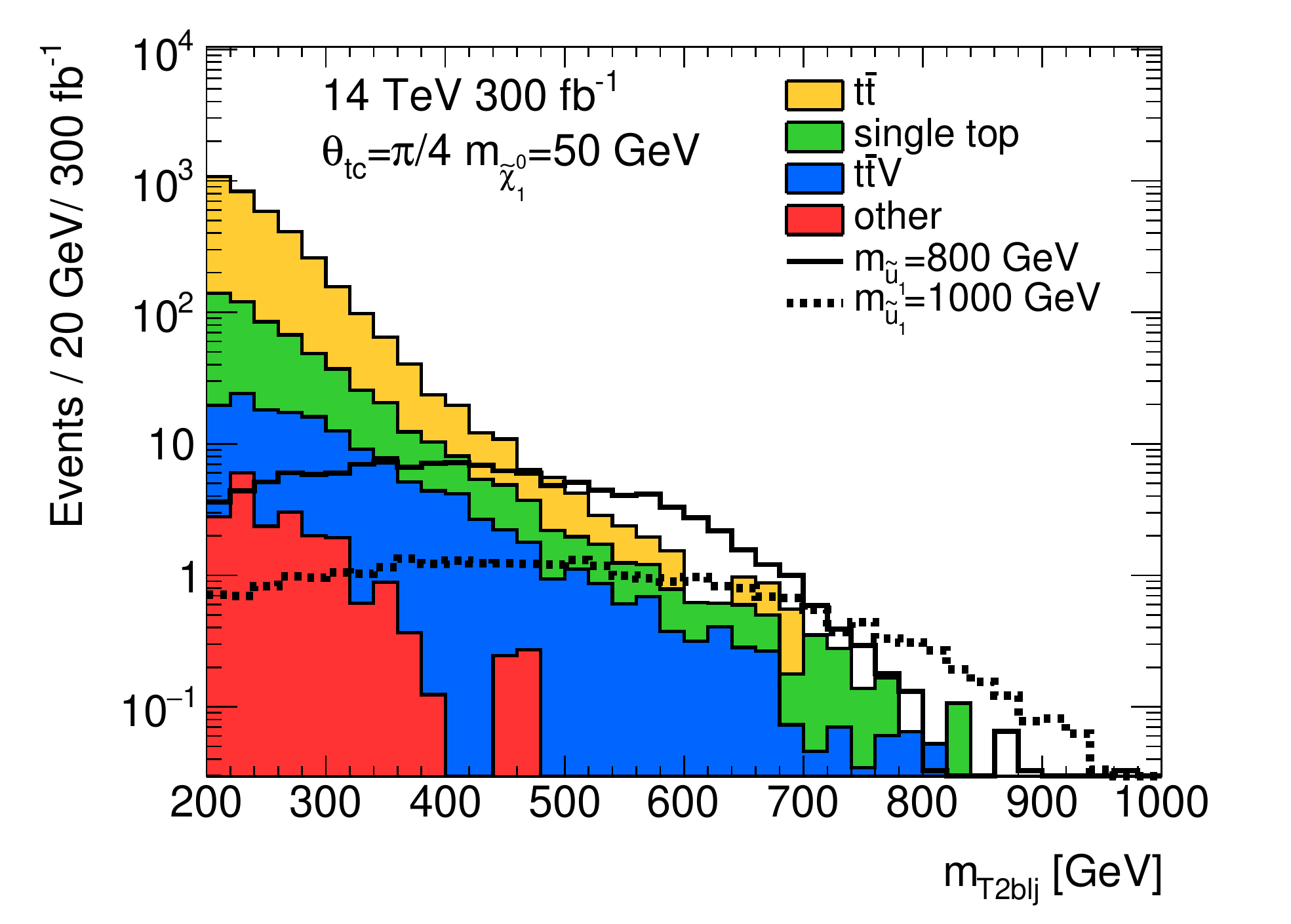}
	\includegraphics[width=0.495\textwidth]{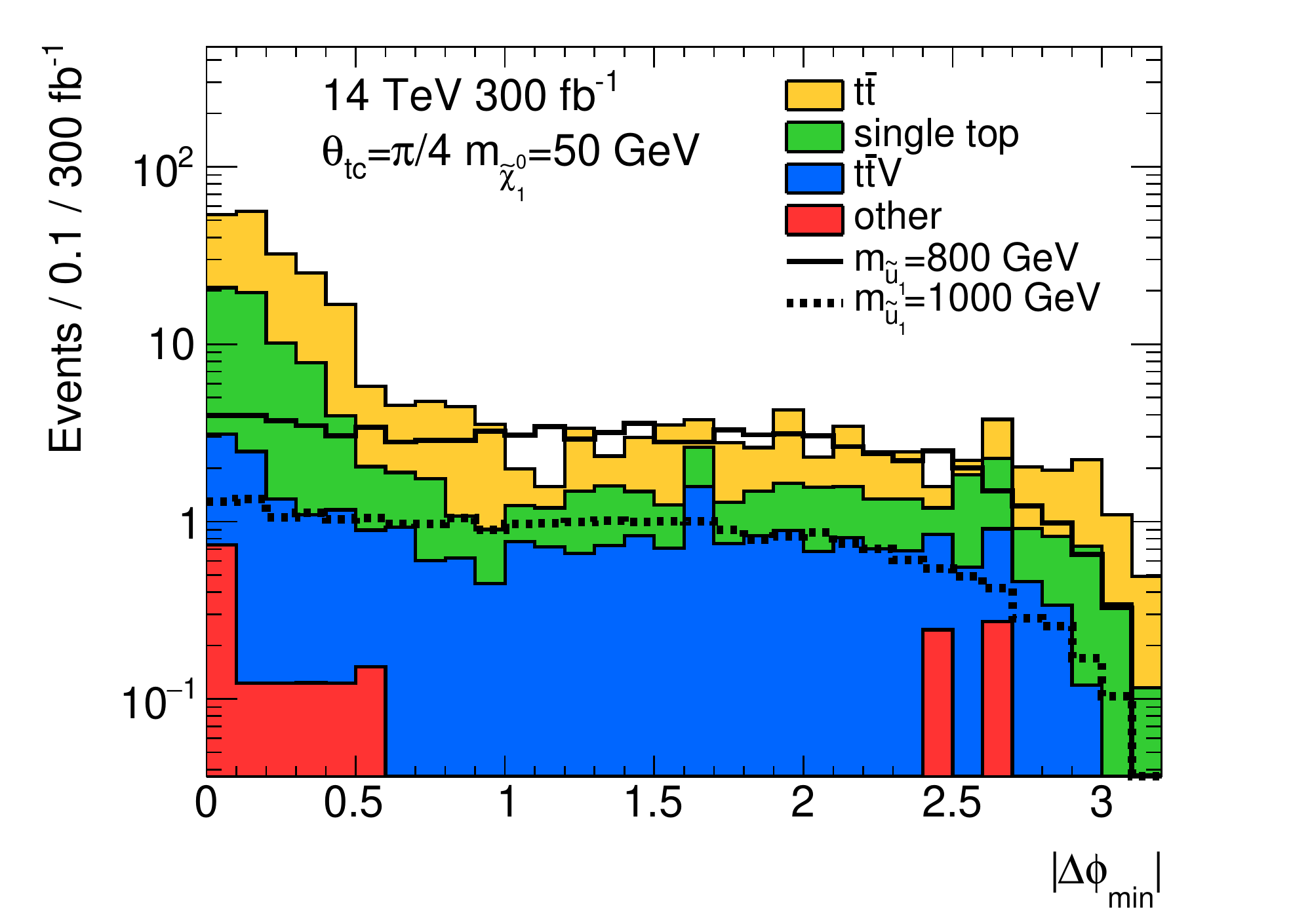}
	\end{center}
	\caption{Distributions in the \mttwoblj (left) and
  $|\Delta\phi_{\rm min}|$ (right) variables after imposing all cuts of the
  Case-A analysis strategy, excepted the one corresponding to the represented
  variable. The different background contributions and two representative signal
  scenarios are shown for an integrated luminosity of 300~fb$^{-1}$. For the
  $|\Delta\phi_{\rm min}|$ distribution, the \mttwoblj variable is required to
  be larger than 400 GeV.}
	\label{fig:nm1}
\end{figure*}

In order to further reduce the dileptonic \ttbar background where one of the
leptons escapes identification, we make use of the now standard asymmetric
\mttwo variable (denoted \amttwo)~\cite{Konar:2009qr, Lester:2014yga} that
consists in a variant of the \mttwo observable. The \amttwo variable is built
from two legs (corresponding to the two decay chains) containing both a visible
part and an invisible part, and it requires two test masses corresponding to the
invisible mass attached with each leg. The visible part of the first leg is
built using the sum of the momenta of the $b$-tagged jet and of the lepton, with
a test mass that is set to zero. The visible part of the
second leg is built from the remaining jet with the highest $b$-tagging weight
and $m_W$ is used as a test mass. The targeted background distribution featuring
an end-point at approximately 160~GeV, we impose an $\amttwo>200\, \GeV$ cut. In
addition, the background can be further reduced by constructing another
transverse $m_{T2}$ variable. The signal topology features one squark leg where
there is a hard $c$-jet, such that the distribution in the transverse mass
built from the transverse momentum of the $c$-jet and the one of the
neutralino exhibits an end-point at $(m_{\tilde u_1}^2-m_{\tilde\chi^0_1}^2)^
{1/2}$. This feature can be exploited by constructing an appropriate 
\mttwoblj variable.
The visible part of the first leg is built from the sum of the momenta of
the $b$-tagged jet and of the lepton, together with a vanishing test mass. 
The visible part of the second leg uses the hardest non-$b$-tagged
jet or the $c$-tagged jet for the Case-A and Case-B strategies respectively, and
again a vanishing test mass. We impose a selection on \mttwoblj depending on the
squark-neutralino mass splitting in order to optimise the sensitivity to the
signal. This optimisation is performed by varying the cut threshold from 300 to
600~GeV in steps of 50~GeV.

Finally, it is found that after all cuts, the missing transverse momentum
is aligned with one of the jets of the event for the backgrounds, whereas 
there is no preferential direction for the signal.
We therefore apply a cut on the minimum azimuthal angle separation between any
jet and the missing transverse momentum, $|\Delta\phi_{\rm min}|>0.6$.

As an illustration, we present, in Fig.~\ref{fig:nm1},
the distribution in the \mttwoblj (left) and $|\Delta\phi_{\rm min}|$
variable (right) for the different SM backgrounds and two representative signal
benchmark points. All selection cuts from the Case-A analysis strategy are
imposed, but the one on the represented variable. On the left panel, we can observe that a
selection of \mttwoblj $> 400$~GeV is sufficient to separate the signal from the
background for the lighter of the chosen benchmark models. 
On the right panel, we show instead the $|\Delta\phi_{\rm min}|$
distribution after including a cut of 400~GeV on \mttwoblj. We can again
observe that a significant improvement of the signal to background ratio 
can be achieved by imposing a $|\Delta\phi_{\rm min}|>0.6$ cut.

\section{Results}
\label{sec:results}

On the basis of the analysis strategy outlined in the previous section, we
estimate the LHC sensitivity to supersymmetric scenarios featuring mixed
stop-scharm states with 300~$\ifb$ and 3000~$\ifb$ of integrated luminosity. 
For the latter configuration,  
we assume no modification in the detector performances for the
high-luminosity LHC. The sensitivity is extracted by means of a test statistics
based on a profiled likelihood ratio, and we make use of the CLs method~\cite{%
Read:2002hq} to obtain 95\% confidence level (CL) exclusion limits. The
statistical analysis is performed with the {\sc RooStat} toolkit \cite{%
Moneta:2010pm} and we assume systematic uncertainties of 20\% and 5\% on the SM
backgrounds and on the signal respectively. The results are presented in
terms of the upper limits, at the 95\% CL, on the ratio of the signal yields
to the corresponding benchmark predictions, denoted as
$\sigma^{\mathrm{excl}}/\sigma^{\mathrm{SUSY}}$. 

\begin{figure*}
	\begin{center}
	\includegraphics[width=0.495\textwidth]{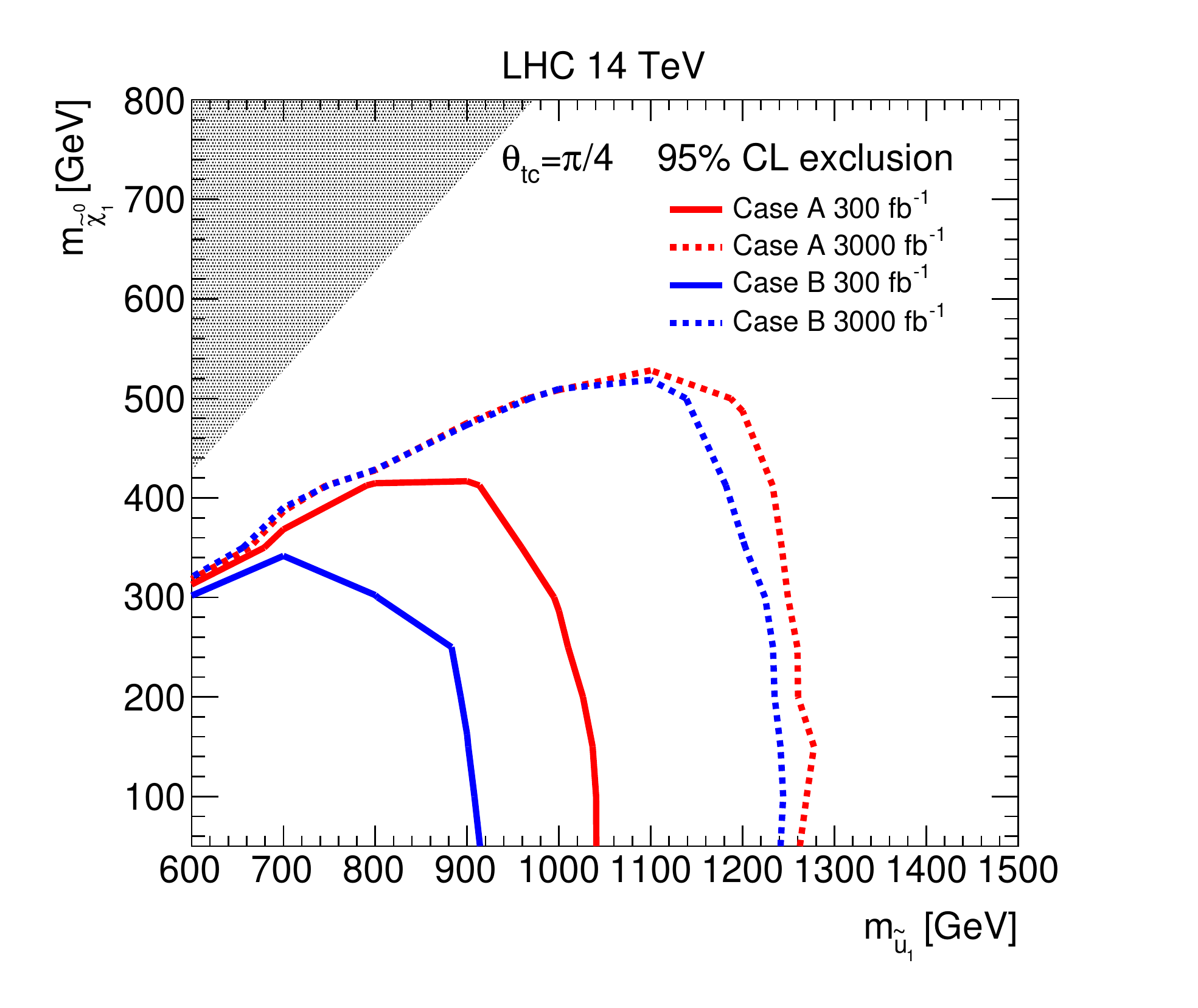}
	\includegraphics[width=0.495\textwidth]{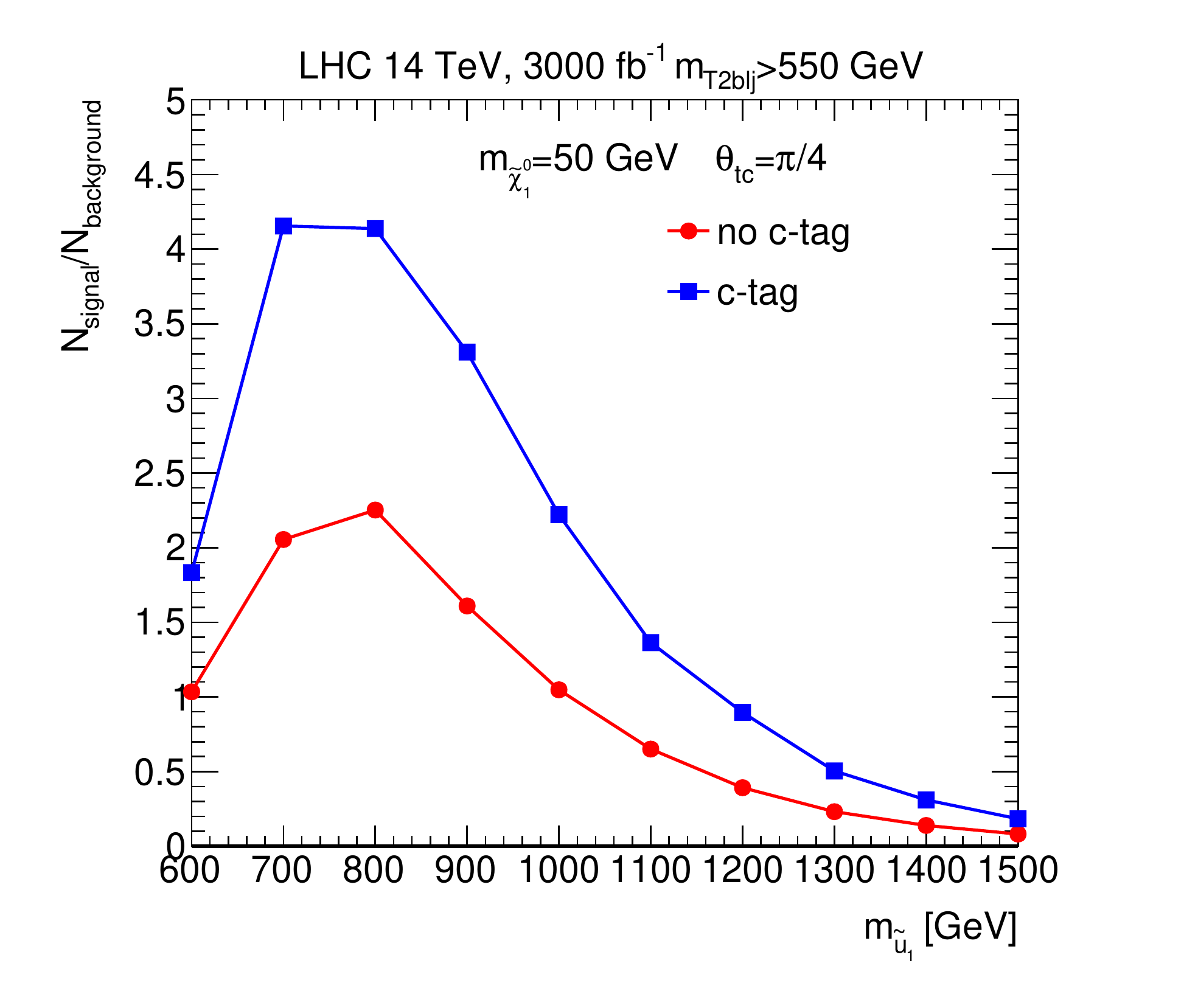}
	\end{center}
	\vspace*{-8mm}
	\caption{Left: Sensitivity of the LHC to our mixed stop-scharm scenarios
  given as 95\% CL exclusion contours in the $(m_{\tilde{u}_1},
  m_{\tilde{\chi}^0_1})$ plane for the Case-A (red) and Case-B (blue) analysis
  strategies and for 300~fb$^{-1}$ (solid) 3000~fb$^{-1}$ (dashed). The projected excluded region lies 
between the exclusion contour and the bottom-left side of the figure. 
  Right: Signal over background ratio as a function of \muone for the Case-A
  (red, rounds) and Case-B (blue, squares) analysis strategies when one imposes
  that $\mttwoblj>550 \, \GeV$. 
}
	\label{fig:reachmass}
\end{figure*}
We show in the left panel of Fig.~\ref{fig:reachmass} the analysis reach in the
$(m_{\tilde{u}_1},m_{\tilde{\chi}^0_1})$ plane both for the Case-A (red) and
Case-B (blue) analysis strategies and for 300~fb$^{-1}$ (solid) and
3000~fb$^{-1}$ (dashed) of integrated luminosity. The region that 
lies between the exclusion contour and the bottom-left side of the figure will be excluded at the future 
runs of LHC. The expected 95\% 
upper limit on \muone for $m_{\tilde{\chi}^0_1}=50\, \GeV$ is $1050 \GeV$ for Case-A, and $920 \GeV$ for Case-B for
an integrated luminosity of 300~fb$^{-1}$. The large difference is
due to the fact that the analysis reach is in this case dominated
by statistics, which is lower for the analysis based on $c$-tagging,
due to the 30\% efficiency of the chosen $c$-tagging working point.
For an integrated luminosity of 3~ab$^{-1}$ the difference is
reduced, with a reach of $1280 \GeV$  for Case-A and $1240 \GeV$ for
Case-B.\par

In order to better understand the relative performance of the two analysis
strategies, we present in the right panel of Fig.~\ref{fig:reachmass} the
dependence of the signal over background ratio ($S/B$) on the squark mass 
$m_{\tilde{u}_1}$ for $m_{\tilde{\chi}^0_1}=50\, \GeV$, $\theta_{tc}=\pi/4$ and
when a \mttwoblj $>$~550~GeV cut is applied. As expected,
the $S/B$ ratio is higher when $c$-tagging is incorporated. 
Comparisons of results stemming from analyses with and
without $c$-tagging, or relying on different $c$-tagging working points could be
used to get information on the flavour content of the observed squark, which is
the main information  one would like to extract
in case of a discovery. In the Case-B analysis, we
have chosen a $c$-tagging working point which optimises $S/B$, 
but with a similar efficiency for $c$-jets and light jets,
and thus not ideal for discriminating the flavour of the signal.
A different $c$-tagging algorithm working point featuring a very high 
rejection for light jets, as {\it e.g.} in Ref.~\cite{Aaboud:2018zjf} 
with $\epsilon_c(c)$=18\%, $1/\epsilon_c(b)=$ 20 and $1/\epsilon_c(l)=200$
would yield a lower overall sensitivity, but might be used to 
discriminate between different flavour mixing hypotheses for the signal. 

In Fig.~\ref{fig:exclu}, we show the 95\% CL exclusion limits in
the $(m_{\tilde u_1}, \theta_{tc})$ plane for a fixed neutralino mass of 50~GeV.
Recasts of the 13~TeV exclusion limits obtained by the ATLAS experiment
with 36 $\ifb$ are in addition overlaid (see Sec.~\ref{sec:recast}), the blue
curve corresponding to the ATLAS search for stop in the singly-leptonic
mode~\cite{Aaboud:2017aeu} and the red one to the ATLAS search for squarks
based on charm-tagging~\cite{Aaboud:2018zjf}. The region between the recasted exclusion contour and 
the top-left (bottom-left) side of the figure is excluded by the $t\bar{t}$+\etmiss 
($c\bar{c}$+\etmiss) analysis. The exclusion limits expected from the $tc$+\etmiss Case-A analysis strategy
are shown as solid and dotted black lines for integrated luminosities of 300
$\ifb$ and 3000 $\ifb$ respectively. Moreover, we also include the expectation
of such an analysis at 13~TeV, with 36~$\ifb$ of luminosity. For $tc$+\etmiss analysis, the 
projected excluded region lies between the exclusion contour (solid, dashed and dotted black lines) 
and the left side of the figure. This figure clearly
illustrates the strength of the $tc$+\etmiss analysis we are proposing, 
covering a region of the parameter space not accessible with current searches 
relying on the MFV paradigm. 

 \begin{figure} 
  \centering 
    \includegraphics[width=\columnwidth]{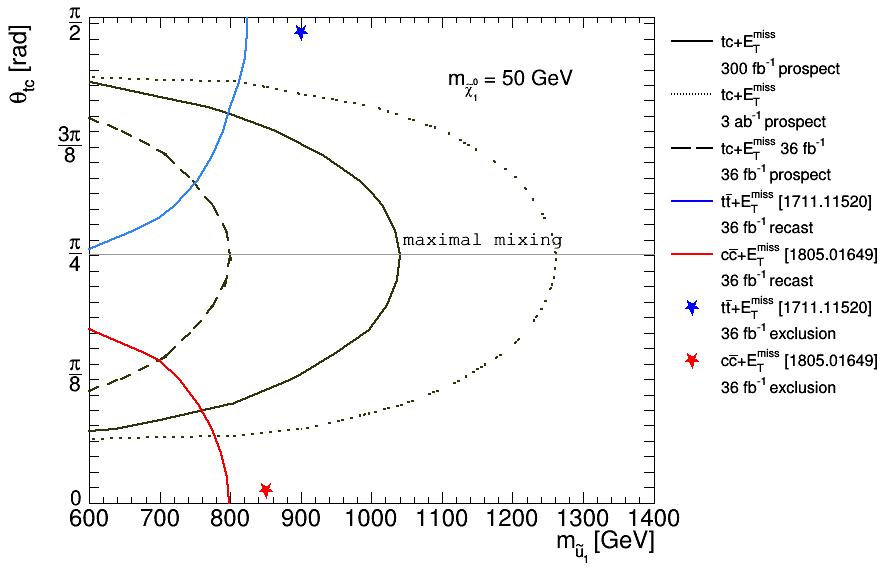}
    \caption{Present and expected exclusion limits in the $(\theta_{tc},
    m_{\tilde u_1})$ plane. The area between the recasted exclusion contour and the top-left (bottom-left) side of 
the figure is excluded by the $t\bar{t}$+\etmiss ($c\bar{c}$+\etmiss) analysis. The region that lies 
between the exclusion contour (solid, dashed and dotted black lines) and the left side of the figure 
will be excluded at the future runs of LHC using 
$tc$+\etmiss analysis. See the text for details.}
\label{fig:exclu}
 \end{figure}


\section{Summary}
\label{sec:summary}

We have studied a supersymmetric scenario departing from the
traditional MFV paradigm. Flavour mixing including squarks of the first
generation and left-handed partners being strongly constrained by data, we focus
on specific scenarios where right-handed stop and scharm mix. In this context,
squark pair production and decay can yield three distinctive signatures, namely
top pairs plus missing energy, charm-jet pairs with missing energy, and $tc$
plus missing energy.

By recasting existing LHC searches for
top and charm squarks, we have shown that a significant fraction of the
parameter space is evading present LHC constraints and that mixed squark
states with masses around 600~GeV remain a phenomenologically viable 
option. This is in particular true when the top-charm mixing is maximal, 
as both $t\bar{t}$ + \etmiss and $c\bar{c}$+\etmiss
signals are weakened. This apparent loophole
in the experimental searches could be filled by the design of
a dedicated  $tc$+\etmiss search for squarks. 

To this purpose,
we have developed two search strategies targeting the production of squarks
containing two flavour components, both involving leptons, $b$-tagged jet and
\etmiss. The core difference between them consists in using charm-tagging 
in one of them, making hence use of the presence of a charm jet in the final 
state as a  handle on new physics. With charm-tagging efficiencies presently 
achieved by experiments being lower than for  
$b$-tagging, the requirement of a charm-tagged jet implies a loss 
of signal statistics and hence of mass reach for an integrated luminosity 
of 300~$\ifb$. The statistical limitation on the signal are 
however less important for the high-luminosity phase of the LHC, 
where mixed squarks as  heavy as 1.3 TeV could be targeted for both 
analysis strategies. 
The advantage of pursuing several analyses
in parallel is that the comparison of the results from 
analyses relying on  different requirement and/or working points 
for the flavour tagging of jets opens the door to probing the squark 
flavour structure and hence allows the potential observation of departures from 
the MFV paradigm.

As charm tagging is being actively developed by both the ATLAS and CMS 
collaborations, we urge them to add to their search program a 
dedicated $tc$+\etmiss analysis that could provide sensitivity  
to new physics scenarios where the top partners are not
flavour eigenstates.

\section*{Acknowledgements}

The authors would like to thank the organizers of 
`Physics at TeV Colliders' workshop (Les Houches, June 2017) 
whe\-re this work was initiated. We would 
also like to thank Michihisa Takeuchi for many useful discussions. This work has 
been partially supported by French state funds
managed by the Agence Nationale de la Recherche (ANR) in the context of the
{\it Investissements d'avenir} Labex ENIGMASS (ANR-11-LABX-0012) and Labex ILP
(ANR-11-IDEX-0004-02, ANR-10-LABX-63), and by the Grant-in-Aid for
Scientific Research on Scientific Research B (No. 16H03991) and Innovative Areas (16H06492).

\bibliographystyle{spphys}
\bibliography{references}

\begin{thebibliography}{10}
\providecommand{\url}[1]{{#1}}
\providecommand{\urlprefix}{URL }
\expandafter\ifx\csname urlstyle\endcsname\relax
  \providecommand{\doi}[1]{DOI \discretionary{}{}{}#1}\else
  \providecommand{\doi}{DOI \discretionary{}{}{}\begingroup
  \urlstyle{rm}\Url}\fi

\bibitem{Sirunyan:2017kqq}
A.M. Sirunyan, et~al., Eur. Phys. J. \textbf{C77}(10), 710 (2017).
\newblock \doi{10.1140/epjc/s10052-017-5267-x}

\bibitem{Aaboud:2017dmy}
M.~Aaboud, et~al., JHEP \textbf{09}, 084 (2017).
\newblock \doi{10.1007/JHEP09(2017)084}

\bibitem{Sirunyan:2017xse}
A.M. Sirunyan, et~al., JHEP \textbf{10}, 019 (2017).
\newblock \doi{10.1007/JHEP10(2017)019}

\bibitem{Sirunyan:2017wif}
A.M. Sirunyan, et~al., JHEP \textbf{10}, 005 (2017).
\newblock \doi{10.1007/JHEP10(2017)005}

\bibitem{Sirunyan:2017kiw}
A.M. Sirunyan, et~al., Phys. Lett. \textbf{B778}, 263 (2018).
\newblock \doi{10.1016/j.physletb.2018.01.012}

\bibitem{Aaboud:2017nfd}
M.~Aaboud, et~al., Eur. Phys. J. \textbf{C77}(12), 898 (2017).
\newblock \doi{10.1140/epjc/s10052-017-5445-x}

\bibitem{Aaboud:2017wqg}
M.~Aaboud, et~al., JHEP \textbf{11}, 195 (2017).
\newblock \doi{10.1007/JHEP11(2017)195}

\bibitem{Aaboud:2017ayj}
M.~Aaboud, et~al., JHEP \textbf{12}, 085 (2017).
\newblock \doi{10.1007/JHEP12(2017)085}

\bibitem{Sirunyan:2017pjw}
A.M. Sirunyan, et~al., Phys. Rev. \textbf{D97}(1), 012007 (2018).
\newblock \doi{10.1103/PhysRevD.97.012007}

\bibitem{Sirunyan:2017leh}
A.M. Sirunyan, et~al., Phys. Rev. \textbf{D97}(3), 032009 (2018).
\newblock \doi{10.1103/PhysRevD.97.032009}

\bibitem{Aaboud:2017phn}
M.~Aaboud, et~al., JHEP \textbf{01}, 126 (2018).
\newblock \doi{10.1007/JHEP01(2018)126}

\bibitem{Aaboud:2017aeu}
M.~Aaboud, et~al., JHEP \textbf{06}, 108 (2018).
\newblock \doi{10.1007/JHEP06(2018)108}

\bibitem{Aaboud:2017vwy}
M.~Aaboud, et~al., Phys. Rev. \textbf{D97}(11), 112001 (2018).
\newblock \doi{10.1103/PhysRevD.97.112001}

\bibitem{Sirunyan:2017cwe}
A.M. Sirunyan, et~al., Phys. Rev. \textbf{D96}(3), 032003 (2017).
\newblock \doi{10.1103/PhysRevD.96.032003}

\bibitem{Aaboud:2018zjf}
M.~Aaboud, et~al., JHEP \textbf{09}, 050 (2018).
\newblock \doi{10.1007/JHEP09(2018)050}

\bibitem{Martin:2007gf}
S.P. Martin, Phys. Rev. \textbf{D75}, 115005 (2007).
\newblock \doi{10.1103/PhysRevD.75.115005}

\bibitem{Fan:2011yu}
J.~Fan, M.~Reece, J.T. Ruderman, JHEP \textbf{11}, 012 (2011).
\newblock \doi{10.1007/JHEP11(2011)012}

\bibitem{Murayama:2012jh}
H.~Murayama, Y.~Nomura, S.~Shirai, K.~Tobioka, Phys. Rev. \textbf{D86}, 115014
  (2012).
\newblock \doi{10.1103/PhysRevD.86.115014}

\bibitem{Blanke:2013zxo}
M.~Blanke, G.F. Giudice, P.~Paradisi, G.~Perez, J.~Zupan, JHEP \textbf{06}, 022
  (2013).
\newblock \doi{10.1007/JHEP06(2013)022}

\bibitem{Brooijmans:2018xbu}
G.~Brooijmans, et~al., in \emph{{10th Les Houches Workshop on Physics at TeV
  Colliders (PhysTeV 2017) Les Houches, France, June 5-23, 2017}} (2018).
\newblock
  \urlprefix\url{http://lss.fnal.gov/archive/2017/conf/fermilab-conf-17-664-ppd.pdf}

\bibitem{Ciuchini:2007ha}
M.~Ciuchini, A.~Masiero, P.~Paradisi, L.~Silvestrini, S.K. Vempati, O.~Vives,
  Nucl. Phys. \textbf{B783}, 112 (2007).
\newblock \doi{10.1016/j.nuclphysb.2007.05.032}

\bibitem{DeCausmaecker:2015yca}
K.~De~Causmaecker, B.~Fuks, B.~Herrmann, F.~Mahmoudi, B.~O'Leary, W.~Porod,
  S.~Sekmen, N.~Strobbe, JHEP \textbf{11}, 125 (2015).
\newblock \doi{10.1007/JHEP11(2015)125}

\bibitem{Dimou:2015yng}
M.~Dimou, S.F. King, C.~Luhn, JHEP \textbf{02}, 118 (2016).
\newblock \doi{10.1007/JHEP02(2016)118}

\bibitem{Dimou:2015cmw}
M.~Dimou, S.F. King, C.~Luhn, Phys. Rev. \textbf{D93}(7), 075026 (2016).
\newblock \doi{10.1103/PhysRevD.93.075026}

\bibitem{Bozzi:2007me}
G.~Bozzi, B.~Fuks, B.~Herrmann, M.~Klasen, Nucl. Phys. \textbf{B787}, 1 (2007).
\newblock \doi{10.1016/j.nuclphysb.2007.05.031}

\bibitem{Fuks:2008ab}
B.~Fuks, B.~Herrmann, M.~Klasen, Nucl. Phys. \textbf{B810}, 266 (2009).
\newblock \doi{10.1016/j.nuclphysb.2008.11.020}

\bibitem{Hurth:2009ke}
T.~Hurth, W.~Porod, JHEP \textbf{08}, 087 (2009).
\newblock \doi{10.1088/1126-6708/2009/08/087}

\bibitem{Bruhnke:2010rh}
M.~Bruhnke, B.~Herrmann, W.~Porod, JHEP \textbf{09}, 006 (2010).
\newblock \doi{10.1007/JHEP09(2010)006}

\bibitem{Bartl:2010du}
A.~Bartl, H.~Eberl, B.~Herrmann, K.~Hidaka, W.~Majerotto, W.~Porod, Phys. Lett.
  \textbf{B698}, 380 (2011).
\newblock \doi{10.1016/j.physletb.2011.04.051, 10.1016/j.physletb.2011.01.020}.
\newblock [Erratum: Phys. Lett.B700,390(2011)]

\bibitem{Bartl:2011wq}
A.~Bartl, H.~Eberl, E.~Ginina, B.~Herrmann, K.~Hidaka, W.~Majerotto, W.~Porod,
  Phys. Rev. \textbf{D84}, 115026 (2011).
\newblock \doi{10.1103/PhysRevD.84.115026}

\bibitem{Bartl:2012tx}
A.~Bartl, H.~Eberl, E.~Ginina, B.~Herrmann, K.~Hidaka, W.~Majerotto, W.~Porod,
  Int. J. Mod. Phys. \textbf{A29}(07), 1450035 (2014).
\newblock \doi{10.1142/S0217751X14500353}

\bibitem{Backovic:2015rwa}
M.~Backovi\'c, A.~Mariotti, M.~Spannowsky, JHEP \textbf{06}, 122 (2015).
\newblock \doi{10.1007/JHEP06(2015)122}

\bibitem{Blanke:2015ulx}
M.~Blanke, B.~Fuks, I.~Galon, G.~Perez, JHEP \textbf{04}, 044 (2016).
\newblock \doi{10.1007/JHEP04(2016)044}

\bibitem{Crivellin:2016rdu}
A.~Crivellin, U.~Haisch, L.C. Tunstall, JHEP \textbf{09}, 080 (2016).
\newblock \doi{10.1007/JHEP09(2016)080}

\bibitem{Blanke:2017tnb}
M.~Blanke, S.~Kast, JHEP \textbf{05}, 162 (2017).
\newblock \doi{10.1007/JHEP05(2017)162}

\bibitem{Blanke:2017fum}
M.~Blanke, S.~Das, S.~Kast, JHEP \textbf{02}, 105 (2018).
\newblock \doi{10.1007/JHEP02(2018)105}

\bibitem{Evans:2015swa}
J.A. Evans, D.~Shih, A.~Thalapillil, JHEP \textbf{07}, 040 (2015).
\newblock \doi{10.1007/JHEP07(2015)040}

\bibitem{Fuks:2014lva}
B.~Fuks, P.~Richardson, A.~Wilcock, Eur. Phys. J. \textbf{C75}(7), 308 (2015).
\newblock \doi{10.1140/epjc/s10052-015-3530-6}

\bibitem{Borschensky:2014cia}
C.~Borschensky, M.~Krämer, A.~Kulesza, M.~Mangano, S.~Padhi, T.~Plehn,
  X.~Portell, Eur. Phys. J. \textbf{C74}(12), 3174 (2014).
\newblock \doi{10.1140/epjc/s10052-014-3174-y}

\bibitem{Alloul:2013bka}
A.~Alloul, N.D. Christensen, C.~Degrande, C.~Duhr, B.~Fuks, Comput. Phys.
  Commun. \textbf{185}, 2250 (2014).
\newblock \doi{10.1016/j.cpc.2014.04.012}

\bibitem{Degrande:2011ua}
C.~Degrande, C.~Duhr, B.~Fuks, D.~Grellscheid, O.~Mattelaer, T.~Reiter, Comput.
  Phys. Commun. \textbf{183}, 1201 (2012).
\newblock \doi{10.1016/j.cpc.2012.01.022}

\bibitem{Alwall:2014hca}
J.~Alwall, R.~Frederix, S.~Frixione, V.~Hirschi, F.~Maltoni, O.~Mattelaer, H.S.
  Shao, T.~Stelzer, P.~Torrielli, M.~Zaro, JHEP \textbf{07}, 079 (2014).
\newblock \doi{10.1007/JHEP07(2014)079}

\bibitem{Ball:2014uwa}
R.D. Ball, et~al., JHEP \textbf{04}, 040 (2015).
\newblock \doi{10.1007/JHEP04(2015)040}

\bibitem{Sjostrand:2014zea}
T.~Sjöstrand, S.~Ask, J.R. Christiansen, R.~Corke, N.~Desai, P.~Ilten,
  S.~Mrenna, S.~Prestel, C.O. Rasmussen, P.Z. Skands, Comput. Phys. Commun.
  \textbf{191}, 159 (2015).
\newblock \doi{10.1016/j.cpc.2015.01.024}

\bibitem{Alioli:2010xd}
S.~Alioli, P.~Nason, C.~Oleari, E.~Re, JHEP \textbf{06}, 043 (2010).
\newblock \doi{10.1007/JHEP06(2010)043}

\bibitem{Lonnblad:2011xx}
L.~Lönnblad, S.~Prestel, JHEP \textbf{03}, 019 (2012).
\newblock \doi{10.1007/JHEP03(2012)019}

\bibitem{Cacciari:2008gp}
M.~Cacciari, G.P. Salam, G.~Soyez, JHEP \textbf{04}, 063 (2008).
\newblock \doi{10.1088/1126-6708/2008/04/063}

\bibitem{Cacciari:2011ma}
M.~Cacciari, G.P. Salam, G.~Soyez, Eur. Phys. J. \textbf{C72}, 1896 (2012).
\newblock \doi{10.1140/epjc/s10052-012-1896-2}

\bibitem{Aad:2008zzm}
{The ATLAS Experiment at the CERN Large Hadron Collider} (2008).
\newblock \doi{10.1088/1748-0221/3/08/S08003}

\bibitem{Aad:2009wy}
{Expected Performance of the ATLAS Experiment - Detector, Trigger and Physics}.
\newblock CERN-OPEN-2008-020 (2009)

\bibitem{Pani:2017qyd}
P.~Pani, G.~Polesello, Phys. Dark Univ. \textbf{21}, 8 (2018).
\newblock \doi{10.1016/j.dark.2018.04.006}

\bibitem{ATL-PHYS-PUB-2015-022}
{Expected performance of the ATLAS $b$-tagging algorithms in Run-2}.
\newblock {ATL-PHYS-PUB-2015-022} (2015).
\newblock \urlprefix\url{https://cds.cern.ch/record/2037697}

\bibitem{ATL-PHYS-PUB-2017-013}
{Optimisation and performance studies of the ATLAS $b$-tagging algorithms for
  the 2017-18 LHC run}.
\newblock {ATL-PHYS-PUB-2017-013} (2017).
\newblock \urlprefix\url{https://cds.cern.ch/record/2273281}

\bibitem{deFavereau:2013fsa}
J.~de~Favereau, C.~Delaere, P.~Demin, A.~Giammanco, V.~Lemaître, A.~Mertens,
  M.~Selvaggi, JHEP \textbf{02}, 057 (2014).
\newblock \doi{10.1007/JHEP02(2014)057}

\bibitem{Konar:2009qr}
P.~Konar, K.~Kong, K.T. Matchev, M.~Park, JHEP \textbf{04}, 086 (2010).
\newblock \doi{10.1007/JHEP04(2010)086}

\bibitem{Lester:2014yga}
C.G. Lester, B.~Nachman, JHEP \textbf{03}, 100 (2015).
\newblock \doi{10.1007/JHEP03(2015)100}

\bibitem{Read:2002hq}
A.L. Read, J. Phys. \textbf{G28}, 2693 (2002).
\newblock \doi{10.1088/0954-3899/28/10/313}.
\newblock [,11(2002)]

\bibitem{Moneta:2010pm}
L.~Moneta, K.~Belasco, K.S. Cranmer, S.~Kreiss, A.~Lazzaro, D.~Piparo,
  G.~Schott, W.~Verkerke, M.~Wolf, PoS \textbf{ACAT2010}, 057 (2010)

\end{thebibliography}

\end{document}